\documentclass[prb,preprint,amsmath,showkeys]{revtex4}

\usepackage{graphicx}    
\usepackage{dcolumn}     
\usepackage{bm}          
\usepackage[T1]{fontenc}
\usepackage {rotating}
\usepackage{lscape}

\begin{document}

\title{First-principles study of ferroelectric oxide epitaxial
       thin films and superlattices:
       role of the mechanical and electrical boundary conditions}

\author{ Javier Junquera }
\affiliation{ Departamento de Ciencias de la Tierra y
              F\'{\i}sica de la Materia Condensada, Universidad de Cantabria,
              Avda. de los Castros s/n, E-39005 Santander, Spain}
\author{Philippe Ghosez}
\affiliation{ Physique Th\'eorique des Mat\'eriaux, Universit\'e de Li\`ege,
              B-4000 Sart Tilman, Belgium}
\date{\today}

\begin{abstract}
 In this review, we propose a summary of the most recent advances
 in the first-principles study of ferroelectric oxide epitaxial thin films 
 and multilayers. We discuss in detail the key roles of mechanical and 
 electrical boundary conditions, providing to the reader the basic 
 background for a simple and intuitive understanding of the evolution 
 of the ferroelectric properties in many nanostructures. Going further
 we also highlight promising new avenues and future challenges within 
 this exciting field or researches.
 
 To appear in "{\it Journal of Computational and Theoretical Nanoscience}"
\end{abstract}


\maketitle


\section{Introduction}

 During the recent years, numerous studies have been reported 
 in the field of ferroelectric oxide nanostructures. These were 
 motivated by the perspective of new technological applications 
 and made possible thanks to combined spectacular developments of 
 experimental and theoretical techniques.\cite{Ahn-04} On the 
 experimental side, it is now possible to grow oxide nanostructures 
 with a control at the atomic scale and to measure their properties 
 using local probes. On the theoretical side, different significant 
 advances have been realized in the field of first-principles calculations 
 that, combined to the recurrent increase of computational power, allow 
 not only to reproduce accurately experimental measurements but also 
 to make trustable predictions that are verified a posteriori. 

 The study of ferroelectric nanostructures is nowadays a broad 
 field of researches and it would be impossible to provide an 
 exhaustive description of earlier works in a short review paper. 
 Our purpose here is to provide a simple and comprehensive 
 description of the origin of ferroelectric finite size effects and to 
 highlight some recent results, focusing essentially on epitaxial thin 
 films and superlattices. 
 For a more complete and detailed description of recent 
 first-principles achievements concerning ferroelectrics and their 
 nanostructures, we refer the reader to recent book chapters 
 \cite{Ghosez-06,Rabe-07,Lichtensteiger-07.2} and topical review 
 papers.\cite{Dawber-05,Rabe-05,Ponomareva-05,Duan-06.2,Scott-06,Setter-06}

 In the present review, we focus our attention on the influence 
 of mechanical and electrical boundary conditions. Although other 
 effects might also be important (chemistry of interfaces, defects ...), 
 both effects play a key role and already allow to understand the main 
 ferroelectric finite size effects when properly taken into account.   
 The roles of mechanical and electrical boundary conditions are 
 introduced separately and their respective influence is first illustrated 
 in the paradigmatic case of a BaTiO$_3$ (the standard ferroelectric 
 perovskite oxide material,  see Sec. \ref{sec:background}) thin film 
 under realistic conditions,  including its epitaxial growth on SrTiO$_3$ 
 (Sec. \ref{sec:strain}), and the screening of its polarization charge 
 by realistic electrodes (Sec. \ref{sec:depfield}). 
 Once the basic physical effects have been introduced,
 the discussion is enlarged to other perovskite oxides epitaxial
 films (Sec. \ref{sec:capacitors}) and superlattices (Sec. \ref{sec:superlattices}). 
 Finally, recent works on other types of ferroelectric oxide nanostructures, 
 such as nanoparticles or nanowires, are briefly reviewed 
 (Sec. \ref{sec:nanoparticles}).

\section{Background}
\label{sec:background}

\begin{figure}[htbp]
 {\par\centering
  {\scalebox{0.38}{\includegraphics{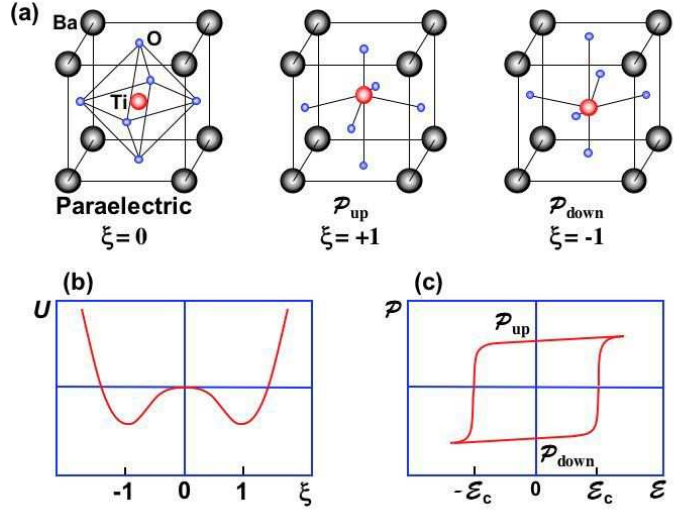}}}
 \par}
 \caption{(a) Crystal structure of BaTiO$_3$ in its high temperature 
          paraelectric cubic perovskite structure and in its room temperature 
          tetragonal structure (for ``up'' and ``down'' polarization states). 
          (b) Typical double-well shape for the internal energy of BaTiO$_3$ 
          in terms of $\xi$ (see text). 
          (c) Hysteretic behavior of the polarization-electric field curve. 
          Figure taken from Ref. \onlinecite{Ghosez-06}.}
\label{Fig.1}
\end{figure}

 Barium titanate (BaTiO$_3$) is a prototypical ferroelectric 
 oxide.\cite{Lines-77} 
 At high temperature, it is paraelectric and crystallizes in the high-symmetry 
 cubic perovskite structure as illustrated in Fig. \ref{Fig.1}a: 
 Ba atom is located 
 at the corner of the cubic unit-cell while the Ti atom is at the center and is
 surrounded by an octahedra of O atoms, themselves located at the center of 
 each face of the cube. On cooling, BaTiO$_3$ undergoes a sequence of three
 ferroelectric phase transitions to structures successively 
 of tetragonal ($T_c \approx 130^{\circ}$C), 
 orthorhombic ($T_c \approx 5^{\circ}$C) and 
 rhombohedral ($T_c \approx -90^{\circ}$C) symmetry. 
 In the following discussion, we will focus on the tetragonal phase 
 that is stable at room temperature. 
 
 The phase transition from cubic to tetragonal is 
 characterized by the opposite shift of Ti and O atoms with respect to Ba, 
 taken as reference. This polar atomic distortion is accompanied with 
 a small relaxation of the unit cell and yields a stable spontaneous 
 polarization of 26 $\mu$C/cm$^2$ at room temperature. 
 In the tetragonal phase, the cubic symmetry is broken, 
 resulting not in a unique polar structure but in six symmetrically 
 equivalent variants with polarization along the [100], [010] and [001] 
 directions.  The two variants along [001] (``up'' and ``down'' states) 
 are represented in Fig. \ref{Fig.1}a. What makes BaTiO$_3$ ferroelectric 
 is not only the existence of energetically equivalent polar states 
 but also the fact that it is possible to switch from one to another 
 of these states by applying an electric field larger than the 
 coercive field. Related to this switchability, the relationship 
 between polarization and electric field is hysteretic in ferroelectrics, 
 as illustrated in Fig. \ref{Fig.1}c. 
 
 In what follows, the pattern of cooperative atomic displacements 
 associated to the phase transition will be referred to as $\xi$, and 
 the paraelectric, up and down states respectively will be labeled 
 as $\xi$ equal to 0, 1 and -1 respectively. 
 For continuous evolution of $\xi$, the internal energy $U$ of the 
 crystal has a typical double-well shape illustrated 
 in Fig. \ref{Fig.1}b. 
 In BaTiO$_3$, the well depth is typically of the order of 30 meV/cell.
 While hysteresis loops (Fig. \ref{Fig.1}c) constitutes the measurement 
 of choice to demonstrate experimentally ferroelectricity, 
 the existence of a double-well shape for the energy is usually 
 considered as the theoretical fingerprint of the ferroelectric instability. 
 Making a Taylor expansion of $U$ in terms of $\xi$ around the 
 paraelectric state taken as reference, and since the odd terms are 
 forbidden by symmetry, we can write

 \begin{eqnarray}
    U(\xi)= A \xi_z^2 + B \xi_z^4 + C \xi_z^6 + D \xi_z^8 + \ldots,
    \label{eq:doublewell}
 \end{eqnarray}

 \noindent where for the sake of simplicity
 we have assumed that atomic displacements 
 in the tetragonal phase can take place only along $z$
 (condition that will be held all along this review unless otherwise stated),
 and we have cut the expansion in the eigth power of the 
 atomic pattern of displacement. 
 $U$ presents a typical double-well shape with a negative 
 curvature at the origin so that, in the previous equation, 
 ferroelectricity can be associated to a negative value of $A$ ($A < 0$). 
 We note that, in the limit of a homogeneous polarization, 
 the previous expansion in terms of microscopic atomic variable $\xi$  
 is similar in spirit to that of Devonshire-Ginzburg-Landau (DGL)
 phenomenological theory \cite{Devonshire-49, Devonshire-51} in terms of the 
 macroscopic polarization, since

 \begin{equation}
    P_z \approx (1/\Omega_0) Z_{zz}^\ast \xi_z,
    \label{eq:pol}
 \end{equation}

 \noindent where $\Omega_0$ is the 
 unit cell volume and $Z_{zz}^\ast$ is the diagonal component of
 $Z^\ast$, the effective charge tensor associated 
 to $\xi$, so that part of the forecoming discussion can be 
 transposed in that alternative context. 
 
 At the bulk level, a coherent microscopic model explaining 
 the origin of ferroelectricity was first reported by 
 Cochran,\cite{Cochran-60} who identified an intimate link between 
 the structural phase transition and the lattice dynamics and 
 introduced the concept of soft mode. Specifically, the underlying 
 idea is that the lowest-frequency  zone-center polar phonon mode 
 in the paraelectric phase becomes softer as a function of 
 decreasing temperature, and finally goes down to zero frequency, 
 freezing in below $T_c$ to generate the ferroelectric crystal structure. 
 Cochran's theory of displacive phase transition was exhibited in the 
 context of a shell-model and the softening of a given polar mode 
 was explained from a competition between short-range and long-range
 Coulomb interactions. 

 Most of these ideas were further confirmed from first-principles 
 calculations \cite{Rabe-07} performed within the density 
 functional theory (DFT). 
 An unstable polar mode
 (i.e. a mode with imaginary frequency $\omega$ within the harmonic 
 approximation) has been identified at $\Gamma$ in the paraelectric phase
 of various ferroelectric oxides \cite{ZhongW-94.1} and, in prototypical 
 cases like BaTiO$_3$, a strong resemblance has been observed between 
 the eigendisplacement vector of this unstable phonon and the 
 ground-state distortion $\xi$ 
 (overlap of 99 \% in BaTiO$_3$ or LiNbO$_3$\cite{Veithen-02.1}). 
 Since energy curvature is related to the square of the frequency, 
 an unstable mode with imaginary frequency in the paraelectric phase 
 is the indication of a negative curvature of the energy at 
 the origin [$A<0$ in Eq. (\ref{eq:doublewell})]. The amplitude of $\omega$ 
 does not directly provide the double-well depth that additionally 
 depends on anharmonic effects but measures, to some extent, the 
 strength of the instability.   The origin of the ferroelectric 
 instability was also investigated at the first-principles 
 level and the competition between short-range (SR) interactions 
 and long-range (LR) Coulomb forces has been confirmed and 
 quantified.\cite{Ghosez-96}  In this context, the anomalous 
 values of the Born effective charges of ferroelectric 
 oxides \cite{ZhongW-94.1,Ghosez-98.2} was shown as an essential feature 
 to produce an unusually large destabilizing Coulomb 
 interaction responsible for the ferroelectric instability.
 
 The balance of forces resulting in the ferroelectric instability is also 
 well-known to be strongly sensitive to strain and external pressure. 
 The sensitivity of the ferroelectric distortion to strain is inherent to the 
 piezoelectric behavior and was highlighted from the early stage of 
 first-principles computations.\cite{Cohen-92,King-Smith-94}  Samara 
 {\it et al.} \cite{Samara-75} proposed that external hydrostatic pressure 
 reduces and eventually totally suppresses ferroelectricity since SR 
 repulsions increases more rapidly than LR destabilizing forces under 
 pressure. This latter result was recently reinvestigated 
 at the first-principles 
 level,\cite{Kornev-05,Bousquet-06} confirming a rapid disappearance of 
 ferroelectricity at moderate pressure and additionally pointing out an 
 unexpected recovery of ferroelectricity at extremely high pressure.
 
 Such microscopic understanding of ferroelectricity at the bulk level 
 allows to anticipate the existence of strong finite size effects 
 in ferroelectrics since both SR and LR forces will be modified 
 in nanostructures. The SR forces will be modified at surface 
 and interfaces due to the change of atomic environment. 
 The same is true for the Born effective charges that cannot 
 keep their bulk value \cite{Ruini-98} as it was illustrated on 
 BaTiO$_3$ thin films by Fu and coworkers.\cite{Fu-99} Beyond that, 
 the LR Coulomb interaction will be cut because of the 
 finite size of the sample and will be strongly dependent 
 on the electrical boundary conditions. Finally, since the ferroelectric 
 instability is strongly sensitive to strains, it will be influenced by 
 mechanical boundary conditions such as epitaxial strains.\cite{Dieguez-04} 
 
 These different factors can act independently to either enhance or 
 suppress ferroelectricity. They will compete with each other 
 in such a way that it is difficult to predict the ferroelectric 
 properties of nanostructures without making explicit investigations. 

\section{Epitaxial thin films}
\label{sec:films}

 During the recent years, most of the efforts devoted to understanding 
 ferroelectric finite size effects concerned ferroelectric thin films.
 This was partly motivated by the observation at the end of the nineties
 of a polar ground-state in ultrathin films \cite{Bune-98, Tybell-99}
 with thicknesses well below 10-20 nm, that was still considered at that
 time as a realistic estimate of critical thickness for 
 ferroelectricity.\cite{Li-97}

 It is now understood that the survival or disappearance of 
 ferroelectricity in ultrathin films is not a truly intrinsic property: 
 it strongly depends on the structure and chemistry of the
 interface between the ferroelectric and the substrate and/or 
 the electrodes as well as on electrical (e.g. screening of  the 
 depolarizing field) and mechanical (e.g. epitaxial strains) 
 boundary conditions. On top of that, additional features such as 
 the finite conductivity of the film or the presence of 
 structural defects and vacancies might also 
 affect the ferroelectric properties. Although these latter features 
 are potentially important, the proper treatment of their effect remains 
 beyond the scope of first-principles simulations and is not yet 
 fully understood so that these effects will not be further discussed here. 
 For an overview of atomic relaxations and discussions of the stability of 
 surfaces and interfaces we refer the reader to previous 
 reviews.\cite{Dawber-05,Ghosez-06,Scott-06} 
 In what follows, we will restrict the discussion on the role of mechanical and 
 electrical the boundary conditions which,  in most cases, provide a sufficient 
 background to achieve relevant understanding of observed ferroelectric finite 
 size effects.

\subsection{Mechanical boundary conditions : the epitaxial strain}
\label{sec:strain}

 As mentioned in Sec. \ref{sec:background}, ferroelectricity in bulk compounds 
 is well-known to be very sensitive to strains.\cite{Lines-77,Cohen-92} 
 In thin films, it is therefore expected that epitaxial strains 
 will play an important role to monitor the ferroelectric properties. 
 Coherently with that, strain engineering of the ferroelectric properties 
 has been recently demonstrated experimentally: 
 a polarization 250\% higher than in bulk single crystals 
 has been reported in BaTiO$_3$ epitaxial films,\cite{Choi-04}
 and room temperature ferroelectricity 
 has even been induced by strain in epitaxial SrTiO$_3$ thin 
 films.\cite{Haeni-04} 

 In order to understand these effects, let us generalize our expansion 
 of the internal energy in terms of additional strain $e_{ij}$ (where
 $i$ and $j$ are cartesian directions) 
 degrees of freedom. 
 In our paradigmatic example of a BaTiO$_3$ film epitaxially grown on a 
 (001) cubic SrTiO$_3$ substrate we have mixed 
 strain/stress boundary conditions: on the one hand the in-plane strains 
 $e_{xx} = e_{yy}$ are fixed by the lattice mismatch between 
 BaTiO$_3$ and SrTiO$_3$, while $e_{xy}=0$; on the other hand, the 
 out-of-plane strain $e_{zz}$ and the 
 shear strains $e_{xz}$ and $e_{yz}$ are free to relax 
 (condition of zero stress: $\sigma_{zz}=\sigma_{xz}=\sigma_{yz}=0$).  
 Again, assuming for simplicity only an homogeneous polarization 
 along $z$-direction, vanishing shear strains, and restricting the expansion to 
 leading orders in $\xi$ and $e$, the free energy functional 
 to be minimized now reads

\begin{eqnarray}
  U(\xi,e) & = & A \xi_z^2 + B \xi_z^4 + C \xi_z^6 + D \xi_z^8 \nonumber \\
           &   & + \frac{1}{2} C_{11} ( 2 e_{xx}^2 + e_{zz}^2 ) 
                 + \frac{1}{2} C_{12} (2e_{xx}^2 + 4 e_{xx} e_{zz}) \nonumber \\
           &   & + 2 g_0 e_{xx} \xi_z^2 + ( g_0 + g_1 ) e_{zz} \xi_z^2.
  \label{eq:Uwithstrain}
\end{eqnarray}

 \noindent The  terms in the first line correspond to the double-well 
 energy of Eq. (\ref{eq:doublewell}). The terms in the second-line are 
 the elastic energy while the terms in the third line arise from the 
 coupling between ionic and strain degrees of freedom. 
 They correspond to the so-called ``polarization-strain coupling'' 
 and are at the origin of the piezoelectric response. 
 It is clear from Eq. (\ref{eq:Uwithstrain}) that 
 the polarization-strain coupling terms
 are responsible for a renormalization of 
 the quadratic part of $U$ that now takes the form

\begin{eqnarray}
   \left[ A + 2 g_0 e_{xx} + \left(g_0+g_1 \right) e_{zz} \right] \xi_z^2. 
   \label{eq:cuadrafterstr}
\end{eqnarray}

\noindent Depending of the value of the parameters $g_0$ and $g_1$,
 and of $e_{xx}$ and $e_{zz}$ 
 (deduced from the relation
 $\partial U/\partial e_{zz}=0$ which follows from the boundary condition
 $\sigma_{zz} = 0$), we see that, 
 playing properly with the epitaxial strain conditions,
 it is possible to modulate the coefficient of $\xi^2$ in order 
 to increase (resp. decrease) its negative value and therefore 
 enhance (resp. suppress) the ferroelectric character of the film. 

\begin{figure}[htbp]
 {\par\centering
  {\scalebox{0.50}{\includegraphics{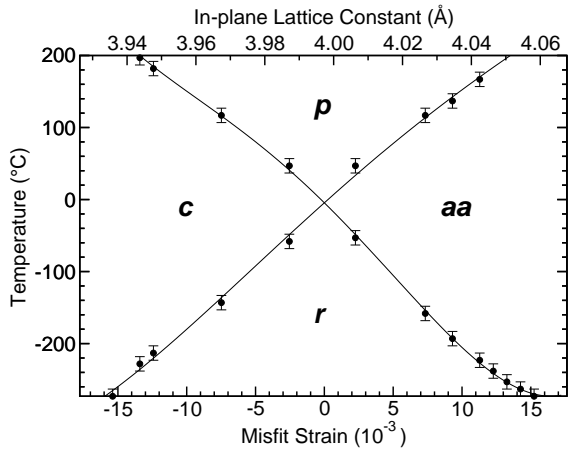}}}
 \par}
\caption{Phase diagram of BaTiO$_3$ in terms of temperature and epitaxial 
         misfit strain. The notation established 
         by Pertsev {\it et al.} \cite{Pertsev-98} is followed:
         $p$ $\equiv$ paraelectric phase; 
         $r$ $\equiv$ monoclinic phase, related with the rhombohedral 
         bulk phase; 
         $c$ $\equiv$ tetragonal phase with an out-of-plane polarization along
         the normal; 
         $aa$ $\equiv$ orthorrombic phase with the polarization along the
         cube face diagonal perpendicular to the normal. 
         Figure taken from Ref. \onlinecite{Dieguez-04}.}
\label{Fig.2}
\end{figure}

 The simplified free energy $U$ in Eq. (\ref{eq:Uwithstrain})
 can be generalized 
 to include other orientations of $\xi$ as well as shear strains and 
 the different parameters can be directly determined from DFT 
 calculations at the bulk level.\cite{King-Smith-94,Dieguez-05} 
 The coefficients of the expansion for a series of ferroelectric perovskites
 have been reported by Di\'eguez {\it et al.} \cite{Dieguez-05}
 These energy parametrizations at 0 K are the starting point
 for the development of model hamiltonians,\cite{ZhongW-94.2,ZhongW-95.1}
 that combined with classical Monte Carlo simulations allow the
 first-principles based studies of temperature versus strain phase diagrams
 in perovskite oxide films, as illustrated for BaTiO$_3$ 
 on a cubic (001) substrate in Fig. \ref{Fig.2}.
 Although effects such as free top surface relaxation or interaction 
 with the substrate are neglected, the phase sequence and the topology 
 of the phase diagram are usually well reproduced within this approach. 
 The predicted phase transitions temperatures are however typically too 
 low compared to those experimentally measured.

 Alternatively, and in fact first applied to perovkskites by Pertsev 
 and coworkers
 \cite{Pertsev-98} before first-principles investigations, similar phase 
 diagram can be obtained from phenomenological 
 Devonshire-Ginzburg-Landau (DGL) type expansion of the free energy 
 in terms of polarization and strain. Results have for instance been 
 reported for  BaTiO$_3$,\cite{Pertsev-98,Pertsev-99,Choi-04}
 PbTiO$_3$,\cite{Pertsev-98} SrTiO$_3$,\cite{Pertsev-00,Pertsev-02,Haeni-04}
 and Pb(Zr$_{x}$Ti$_{1-x}$)O$_{3}$ (PZT) solid solution.\cite{Pertsev-03}
 This provides more direct access to the temperature behavior and 
 gives very accurate results around the temperature/strain regime 
 in which the model parameters were fitted (usually near the 
 bulk ferroelectric transition). 
 In distant strain-temperatures regimes their predictions might be less
 accurate, however. 
 For large strains or low temperatures the uncertainty 
 increases and, as reported for BaTiO$_3$, different sets of DGL 
 parameters can provide phase diagrams qualitatively different far 
 from the bulk transition region.\cite{Pertsev-98,Pertsev-99}

 As a general rule for the usual perovskites on a (001) 
 substrate,\cite{Dieguez-04} sufficiently large epitaxial compressive 
 (resp. tensile) strains will favor a ferroelectric $c$-phase 
 (resp. $aa$-phase) with  out-of-plane (resp. in-plane) 
 polarization while distinct behaviors are predicted for the 
 different compounds in the intermediate regime. 
 Of course, there might be some discrepancies with this rule,
 such as the one pointed by Catalan and coworkers \cite{Catalan-06}
 on PbTiO$_3$ films under tensile strains.
 
 For the BaTiO$_3$ film on a (001) SrTiO$_3$ substrate previously considered, 
 the large compressive epitaxial strain (-2.38 \%) will strongly 
 increase the ferroelectric phase transition temperature and favor a 
 tetragonal $c$-phase with an out-of-plane polarization.
 As further estimated by Neaton and Rabe,\cite{Neaton-03,Tian-06} 
 such epitaxial contraint should enhance the polarization 
 at 0 K by more than 50 \%.  The sensitivity of polarization to 
 strain  is however not expected to be always so large:  in highly polar 
 Pb-based systems such as Pb(Zr$_{x}$Ti$_{1-x})$O$_{3}$ and PbTiO$_3$ 
 where the ferroelectric displacements are already large,  it was reported 
 to be much more reduced than in BaTiO$_3$,\cite{Lee-07} similarly 
 to what was also discussed for BiFeO$_3$ and  
 LiNbO$_3$.\cite{Ederer-05,Ederer-05b}

 In Ref. \onlinecite{Dieguez-04}, like in many studies, the phase diagrams 
 were obtained restricting the investigation to the effect of the 
 epitaxial strain on the ferroelectric degree of freedom $\xi$, or equivalently 
 $P$. In systems where different kinds of instabilites coexist
 (such as ferroelectric and antiferrodistortive instabilities in 
 SrTiO$_{3}$,\cite{ZhongW-95} or the interplay between in-plane 
 ferroelectricity, antiferroelectric and antiferrodistortive distortions
 in PbTiO$_{3}$ free standing slabs \cite{Umeno-06.2})
 the epitaxial strain can modify the competition between these 
 instabilities so that all of them must be explicitely considered and 
 generate much more complicated phase diagrams.\cite{Koukhar-01} 
  Also, up to now, DFT calculations were restricted to 
 mono-domain configurations while much complex domain structures can 
 appear as highlighted using a DGL approach \cite{Koukhar-01b,Emelyanov-01} 
 or an effective Hamiltonian approach \cite{Ponomareva-05,Paul-07} as 
 it will be further illustrated in Section \ref{sec:capacitors}. 

 The effects imposed by the substrate on the lattice constant and 
 geometry of the thin films can be maintained only up to a critical thickness,
 $h_{c}$, above which the elastic deformation can not be maintained any longer.
 Above $h_{c}$ the strain energy is relaxed by the formation of misfit
 dislocations at the film-substrate interface. An estimate of
 $h_{c}$ can be obtained using the Matthews-Blakeslee 
 formula.\cite{Matthews-74} A priori, this structural limit imposes some
 restriction on the potential enhancement of the ferroelectric properties 
 by the mechanical boundary conditions. However, it has been found that 
 a coherent strain state can be preserved well above the theoretical critical 
 thicknesses expected for strain relaxation.\cite{Choi-04}
 Moreover, in some cases, it was reported that the ferroelectric transition 
 temperature still has a value much larger than in bulk, even after the relaxation 
 of the strain state.\cite{Gariglio-07}
 
 In all previous approaches, only the presence of an homogeneous
 strain was considered. However, ``flexoelectric'' coupling
 of a strain gradient to the polarization,\cite{Catalan-05,Ma-06}
 and the related coupling 
 of strain to a polarization gradient is receiving increasing attention.
 Although the characteristic scale of the flexoelectric 
 coefficient is extremely small, it might affect the ferroelectric
 transitions in thin films.
 The subtleties of defining flexoelectric coefficients as
 a bulk property \cite{Tagantsev-86,Tagantsev-85,Resta-90,Klic-04}
 and its computation in a first-principles framework is a challenge for
 theoreticians.

 A more exhaustive  discussion of strain effects can be found in a recent 
 topical review by K. M. Rabe.\cite{Rabe-05}

\subsection{Electrical boundary conditions : the depolarizing field}
\label{sec:depfield}

 As it was explained in Sec. \ref{sec:background}, ferroelectricity 
 in perovskite oxides at the bulk level is a collective
 effect resulting from a delicate balance between SR interactions,
 that favour a paraelectric phase, and LR dipole-dipole electrostatic
 interactions, that favour a ferroelectric distortion.
 So, it is expected that cutting these LR interactions by reducing the
 number of dipoles to interact with in a finite system,
 will affect the ferroelectric behavior of the material.

 If we try to induce a homogeneous out-of-plane polarization in
 a (001) free standing BaTiO$_3$ slab in vacuum, 
 preservation of the normal component of the electric displacement field 
 at the interface will induce a depolarizing field 
 $\mathcal{E}_d = - P/\varepsilon_0$, where $\varepsilon_0$
 is the permittivity of vacuum (unless otherwise stated we will use 
 throughout this review the 
 SI system of units). 
 This depolarizing field totally suppresses the possibility of 
 stabilizing an homogeneous monodomain configuration
 with an out-of-plane polarization.
 
 Many applications rely on films with a uniform switchable out-of-plane
 polarization. So, after the 
 previous discussion, it is clear that, to stabilize such a configuration,
 we need to screen the depolarizing field $\mathcal{E}_{d}$.
 This can be achieved through two basic mechanisms: 
 the first one is the compensation by ``free'' charges provided either by 
 a metallic electrodes, semiconducting substrates, or even by
 ionic adsorbates\cite{Fong-06,Spanier-06}  
 (including in this category the screening
 by external applied fields that might be generated by the free charges or
 dipole layers);\cite{Meyer-01}
 the second one, and the only one if the substrate is an
 insulator,\cite{Streiffer-02,Fong-04}
 is the breaking up of the system into 180$^{\circ}$ domains.
 Most of the fully first-principles calculations have been devoted
 to the screening by free charges, especially due to the very demanding
 computational effort that an explicit treatment of the domains implies,
 so that only the first issue will be discussed in more details here.

 In an early first-principles paper, 
 Ghosez and Rabe,\cite{Ghosez-00.1} using a model Hamiltonian 
 approach, showed that ferroelectricity might be preserved in ferroelectric 
 slabs if $\mathcal{E}_{d}$ is perfectly compensated. 
 This was further confirmed by Meyer and Vanderbilt, at least for
 some terminations of the free standing slabs, in a fully 
 first-principles atomistic calculation.\cite{Meyer-01} 

\begin{figure}[htbp]
 {\par\centering
  {\scalebox{0.50}{\includegraphics{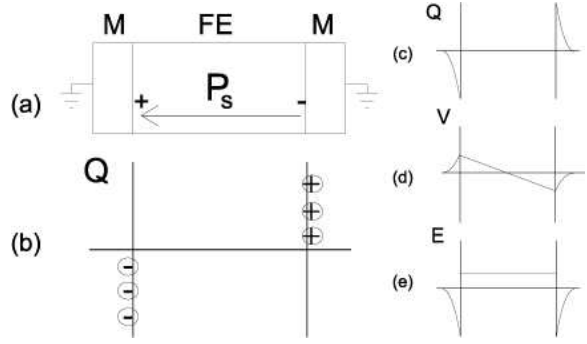}}}
 \par}
 \caption{ A schematic diagram of (a) a short-circuited electrode-ferroelectric
           structure with the spontaneous polarization displayed; 
           (b) the screening charge distribution in the 
           presence of perfect electrodes;
           its (c) charge distribution,
           (d) voltage and
           (e) field profiles in the presence of realistic electrodes.
           Please, note that here the film is taken to be a perfect insulator.
           Figure taken from Ref. \onlinecite{Dawber-03}.}
\label{fig:depolfield}
\end{figure}

 In principle, this total compensation of the depolarizing field 
 could be achieved introducing the ferroelectric film between 
 two metallic electrodes in short-circuit.  
 However, following an idea introduced in the 
 early seventies by Batra,\cite{Batra-72,Wurfel-73,Batra-73}
 Metha \cite{Mehta-73} and coworkers 
 in the context of a phenomenological model (see Fig. \ref{fig:depolfield}), 
 Junquera and Ghosez \cite{Junquera-03.1} demonstrated from 
 first-principles that, even in that case, compensation of the 
 depolarizing field is incomplete, due to the finite screening-length at
 the metal/ferroelectric interface.

 Within this model, if a BaTiO$_3$ film gets a polarization normal to the
 ferroelectric/electrode interface, then the discontinuity of the
 polarization at the interface gives rise to a net interface charge density
 $\sigma_{pol}$ \cite{Feynman} (Fig. \ref{fig:depolfield}a). 
 In order to compensate for this polarization charge density,
 a screening charge is induced in the electrode
 that spreads in practice over a finite distance from the interface
 (Fig. \ref{fig:depolfield}c), 
 giving rise to an interface dipole moment density and a 
 jump in the electrostatic potential at each ferroelectric/electrode interface
 (Fig. \ref{fig:depolfield}d). 
 This potential drop $\Delta V$ is shown 
 to be proportional to the polarization \cite{Junquera-03.1}

\begin{equation}
   \Delta V = \frac{\lambda_{eff}}{\varepsilon_0} P, 
   \label{eq:deltaV}
\end{equation}

 \noindent where $\lambda_{eff}$ is a constant 
 of proportionality hereafter referred to as the effective screening length. 
 Assuming a capacitor with two similar metal/ferroelectric interfaces, 
 the total potential drop into the electrode/ferroelectric/electrode
 heterostructure amounts to $2 \Delta V$. If the electrodes are in short-circuit
 (i.e. at the same potential), a residual field within the ferroelectric must 
 compensate for this drop of potential (Fig. \ref{fig:depolfield}e),

\begin{equation}
  \mathcal{E}_d = -\frac{ 2 \Delta V}{m a_0}, 
  \label{eq:ed1} 
\end{equation}

 \noindent where $m a_0$ is the film thickness in terms of the number $m$ 
 of unit cells of the ferroelectric perovskite oxide.
 Inserting now Eq. (\ref{eq:deltaV}) into Eq. (\ref{eq:ed1})

\begin{eqnarray}
  \mathcal{E}_d = -\frac{2 \lambda_{eff} P}{\varepsilon_0 m a_0}.
  \label{eq:depolfield}
\end{eqnarray}

 \noindent This shows that the field produced by the screening charge 
 within the metal does not fully compensate for the bare depolarizing 
 field $\mathcal{E}_d = - P/\varepsilon_0$: a residual depolarizing 
 field remains that is proportional to $P$ and to the ratio between 
 the effective screening length and the film thickness. It will tend 
 to zero for thick films ($m \rightarrow \infty$) but can become 
 sizable when the film thickness becomes comparable 
 to $\lambda_{eff}$.  The ratio between the residual field into the 
 ferroelectric (Eq. \ref{eq:depolfield}) and the bare depolarizing field
 quantifies the quality of the screening and can be expressed as
 $1-\beta = 2 \lambda_{eff}/ m a_0$ (as it will be further discussed 
 in Sec. III-C). 
 
 The depolarizing field can be computed 
 averaging \cite{Baldereschi-88,Colombo-91,Peressi-98} 
 the microscopic electrostatic 
 potential provided by a first-principles calculation for the whole capacitor.
 Details on the methodology and practical recipes to perform this 
 nano-smoothing can be found in Ref. \onlinecite{Junquera-07}.
 Since the atomic positions, and therefore the induced polarization 
 [Eq. (\ref{eq:pol})], and the thickness of the film are input variables
 of the simulation,  $\lambda_{eff}$  can be estimated using 
 Eq. (\ref{eq:depolfield}) once $\mathcal{E}_{d}$ is known.

 Junquera and Ghosez reported a value of $\lambda_{eff} = 0.23$ \AA\
 (one order of magnitude smaller than the interplanar distances in 
 the heterostructure) for their frozen phonon calculations on
 a SrRuO$_3$/BaTiO$_3$/SrRuO$_3$ capacitor. It might be even slightly 
 smaller when the atomic relaxation of the metal is properly 
 treated.\cite{Gerra-06} 
 It is worth noticing that the effective screening length introduced 
 in this model is NOT an intrinsic  property of the metal 
 (like the Thomas-Fermi screening length), but a quantity that 
 strongly depends also 
 on the details of the ferroelectric/electrode interface such as 
 the different chemical bonds formed 
 at the junction, or the penetration of electronic wave functions of the 
 metal into the ferroelectric and vice versa.
 As shown by Sai {\it et al.},\cite{Sai-05} $\lambda_{eff}$ 
 can vary and become 
 very small for some specific metal/ferroelectric interfaces. 
 Understanding and controlling this compensation mechanism of the 
 polarization charges in order to stabilize a uniform switchable polarization
 in the film is one of the challenges in the field of thin film 
 ferroelectrics.\cite{Dawber-05}

 Other authors \cite{Bratkovsky-00,Bratkovsky-01} alternatively explain the 
 origin of a residual depolarizing field inside the ferroelectric 
 thin film from the 
 presence of a ``dead layer'' at the metal/ferroelectric interface. 
 Although evoking a different mechanism, the theory of the dead layer 
 yields very similar mathematical expressions than the theory of the 
 imperfect metal screening discussed above,  the role of the effective 
 screening length being played by the thickness of the dead layer 
 (see for instance Ref. \onlinecite{Ghosez-06}, or compare
 the formulas of Ref. \onlinecite{Bratkovsky-06.2} by Bratkovsky and Levanyuk 
 with those of Ref. \onlinecite{Chensky-82} by Chensky and Tarasenko). 
 Stengel and Spaldin \cite{Stengel-06} recently reconciled these 
 two apparently antagonist theories, providing a unified interpretation. 
 Using a first-principles approach, they showed that the origin of an 
 intrinsic dead layer with poor dielectric properties at the interface 
 in oxide nanocapacitors is related to the hardening of the collective 
 zone center polar modes that itself is produced by the incomplete 
 screening of the depolarizing field.

 In the presence of a non-vanishing electric field, the proper energy 
 to be considered and minimized is no more the internal energy $U(\xi)$ 
 of the crystal but a field-dependent energy 
 potential $F \left( \xi,\mathcal{E} \right)$ 
 that, on top of $U(\xi)$,  also includes an electrostatic energy 
 term arising from the coupling between the polarization 
 and the field.\cite{Sai-02} This coupling term is
 proportional to $ -\mathcal{E}_{d} \cdot P$. Since the 
 depolarizing field  is 
 proportional to $\left( - P \right)$ [Eq. (\ref{eq:depolfield})] 
 the additional electrostatic term 
 in $F$ is proportional to $P^{2}$ and positive in sign. 
 Remembering Eq. (\ref{eq:pol}),
 we see that proper treatment of the electrostatic energy will 
 produce an additional term in  Eq. (\ref{eq:cuadrafterstr}) 
 of the form $+ \alpha \xi^2$. So, it clearly appears that, in a 
 way similar to the epitaxial strain in Sec. \ref{sec:strain},  
 the incomplete screening of the depolarizing field is responsible 
 for a renormalization of the quadratic coefficient of the energy. 
 In this case however, the correction term is always
 positive ($\alpha \geq 0$) and therefore its effect contributes 
 to suppress ferroelectricity. 

 The depolarizing field is inversely proportional to the thickness
 of the ferroelectric layer, Eq. (\ref{eq:depolfield}).
 The thinnest the film, the larger the depolarizing field and the
 larger the energy penalty to stabilize a monodomain 
 pattern of the polarization.
 In order to minimize the energy while remaining in a homogeneous 
 configuration, the ferroelectric suffers a progressive reduction
 of the polarization that show itself well above the critical thickness.  
 This point was first confirmed experimentally by 
 Lichtensteiger {\it et al.} \cite{Lichtensteiger-05} for PbTiO$_3$ 
 monodomain epitaxial films through X-ray diffraction (XRD) measurement 
 of the evolution of tetragonality $c/a$ with thickness. 
 The tetragonality indeed 
 provides indirect but easily accessible information on the 
 polarization through 
 the polarization-strain coupling 
 [from Eq. (3), $(c/a)_P = (c/a)_0 + \gamma P^2$] 
 so that its measurement has become a standard technique to 
 investigate size 
 effects in thin films 
 and superlattices.\cite{Lichtensteiger-05,Dawber-05.2,Dawber-U}
 The reduction of the polarization was further confirmed by polarization data of
 Kim {\it et al.} \cite{YSKim-05} in ultrathin SrRuO$_3$/BaTiO$_3$/SrRuO$_3$
 capacitors free from passive layers,\cite{YSKim-06} and more 
 recent X-ray photoelectron diffraction (XPD) experiments of 
 Despont {\it et al.}\cite{Despont-06}

 As it was mentioned before, an alternative way of reducing the
 electrostatic energy is the formation of 180$^\circ$ domains.
 In other words, going below the critical thickness predicted for 
 monodomain configuration in ferroelectrics does not necessarily 
 imply a  paraelectric state.\cite{Scott-06}
 The existence of 180$^\circ$ domains prevent the appearance of 
 a {\it net} interfacial charge density.\cite{Speck-94.1}
 This situation is well known in ferroelectric films 
 on insulating substrates, where alternating 180$^\circ$ stripe domains 
 structures have been observed and characterized analyzing the satellite
 peaks in synchrotron X-ray diffraction 
 measurements.\cite{Streiffer-02,Fong-04}
 Even on conducting substrates, where the free charges of the
 electrode should provide a substantial amount of screening,
 the formation of 180$^\circ$ domains has been proposed
 to explain concomitant suppression of polarization and recovery of 
 full tetragonality in Pb(Zr$_{0.2}$Ti$_{0.8}$)O$_3$ on 
 SrRuO$_3$.\cite{Nagarajan-06}
 This is supported with recent works based on the Landau 
 theory,\cite{Bratkovsky-06.1} that conclude that domains always form in
 thin films, almost irrespective of the nature of the electrode
 and whether or not the screening carriers may be present in the
 ferroelectric itself.

 The situation is more complex experimentally,
 and whether the system breaks up into domains or not
 seems to be a very subtle issue.
 Lichtensteiger {\it et al.}, using the same experimental setup,
 have observed how high-quality ultrathin films of PbTiO$_3$
 grown on Nb-SrTiO$_3$ electrodes remain in a
 monodomain configuration \cite{Lichtensteiger-05}
 (although with reduced polarization and tetragonality)
 whereas they form domains when the electrode
 is replaced by
 La$_{0.67}$Sr$_{0.33}$MnO$_3$/SrTiO$_3$.\cite{Lichtensteiger-07}
 Even in the case of insulating substrates,
 the formation of domains depends on how quickly the 
 sample is cooled. 
 Monodomain phases of PbTiO$_3$ on insulating SrTiO$_3$
 have been grown and the structure have been analyzed 
 with a coherent Bragg rod analysis
 (COBRA) method,\cite{Fong-05} when the samples were cooled from deposition
 temperature to room temperature over a period of 24 hours,
 in contrast to the 180$^\circ$ stripe periods observed 
 before.\cite{Streiffer-02,Fong-04}
 Nevertheless, even in the latter case, 
 the system can no more be properly called ferroelectric since, 
 although locally polarized, its polarization is no 
 more expected to be switchable.\cite{Nagarajan-06}

\subsection{ First-principles modeling of ferroelectric capacitors 
             considering mechanical and electrical boundary conditions.}
\label{sec:capacitors}
            
 From  the previous discussion, it is clear that appropriate treatments of 
 the strain and of the incomplete screening of $\mathcal{E}_{d}$
 are crucial to describe size evolution of the ferroelectric properties. 
 Very recently, many DFT calculations as well as first-principles-based
 effective Hamiltonian or atomistic shell-models simulations have
 been reported addressing the problem of the thickness dependence
 of the ferroelectric properties in thin films, and the eventual existence
 of a critical thickness for ferroelectricity.

 \begin{table}
    \caption[ ]{ Review of the most recent fully first-principles simulations
                 on the ferroelectric properties of ferroelectric
                 ultrathin films in a monodomain configuration.
                 Both computations within the local density approximation
                 (LDA) and the generalized gradient approximation (GGA)
                 to the density functional theory have been reported in the
                 literature.
                 CA functional refers to Ceperley-Alder \cite{Ceperley-80}
                 in the parametrization of Perdew-Zunger,\cite{Perdew-81}
                 while PW91 and PBE stand for the 
                 GGA functionals of Perdew and Wang \cite{PW91.1,PW91.2} and
                 Perdew, Burke and Ernzerhof \cite{Perdew-96,Perdew-97} 
                 respectively.
                 Regarding the method,
                 NAO stands for numerical atomic orbitals as implemented
                 in the {\sc Siesta} code,\cite{Soler-02}
                 PW is the acronym for plane waves (as
		 implemented in VASP \cite{Kresse-93,Kresse-96} in 
		 Refs. \onlinecite{Gerra-06,Duan-06}, or 
                 DACAPO \cite{DACAPO}  in Refs. \onlinecite{Sai-05,Fong-06}),
                 and MBPP stands for mixed-basis pseudopotentials.\cite{MBPP}
                 $a_{\parallel}$ represents the in-plane lattice constant.
                 If a theoretical in-plane lattice constant is used to fix the
                 length of the in-plane lattice vectors of the heterostructure,
                 it is computed within the same method and approximations
                 than in the simulation of the whole heterostructure.
                 Different methodologies yield to slightly different values of
                 the lattice constant.
                 $t_{c}$ stands for the critical thickness (if any) of the
                 perovskite ultrathin film, in units of the number of cells 
                 of the ferroelectric perovskite oxide.
               }
    \begin{center}
       {\scriptsize
       \begin{tabular}{|c|c|c|c|c|c|c|}
          \hline
          \hline
           Reference                       &
           Heterostructure                 &
           Method                          &
           Functional                      &
           Interface                       &
           $a_{\parallel}$                 &
           $\rm t_{c}$                     \\
          \hline
          Junquera {\it et al.} [\onlinecite{Junquera-03.1}] &
          SrRuO$_3$/BaTiO$_3$/SrRuO$_3$                      &
          NAO                                                &
          LDA (CA)                                           &
          SrO-TiO$_2$                                        &
          3.874 \AA\ ($a^{\rm th}_{\rm SrTiO_3}$)            &
          6                                                  \\
          Junquera {\it et al.}                              &
          SrRuO$_3$/PbTiO$_3$/SrRuO$_3$                      &
          NAO                                                &
          LDA (CA)                                           &
          SrO-TiO$_2$                                        &
          3.874 \AA\ ($a^{\rm th}_{\rm SrTiO_3}$)            &
          6                                                  \\
          Gerra {\it et al.} [\onlinecite{Gerra-06}]         &
          SrRuO$_3$/BaTiO$_3$/SrRuO$_3$                      &
          PW                                                 &
          GGA (PW91)                                         &
          SrO-TiO$_2$                                        &
          3.94 \AA\ ($a^{\rm th}_{\rm SrTiO_3}$)             &
          3                                                  \\
          Umeno {\it et al.} [\onlinecite{Umeno-06.1}]       &
          Pt/PbTiO$_3$/Pt                                    &
          MBPP                                               &
          LDA (CA)                                           &
          Pt-PbO                                             &
          3.845 \AA\ ($a^{\rm th}_{\rm SrTiO_3}$)            &
          4                                                  \\
                                                             &
                                                             &
                                                             &
                                                             &
          Pt-TiO$_{2}$                                       &
                                                             &
          6                                                  \\
                                                             &
                                                             &
                                                             &
          GGA (PW91)                                         &
          Pt-PbO                                             &
          3.905 \AA\ ($a^{\rm exp}_{\rm PbTiO_3}$)           &
          No                                                 \\
                                                             &
                                                             &
                                                             &
                                                             &
          Pt-TiO$_{2}$                                       &
                                                             &
          No                                                 \\
          Duan  {\it et al.} [\onlinecite{Duan-06}]          &
          SrRuO$_3$/KNbO$_3$/SrRuO$_3$                       &
          PW                                                 &
          LDA (CA)                                           &
          SrO-NbO$_2$                                        &
          3.905 \AA\ ($a^{\rm exp}_{\rm SrTiO_3}$)           &
          4                                                  \\
                                                             &
          Pt/KNbO$_3$/Pt                                     &
                                                             &
                                                             &
          Pt-NbO$_2$                                         &
                                                             &
          2                                                  \\
          Na Sai {\it et al.} [\onlinecite{Sai-05},\onlinecite{Sai-06}] &
          SrRuO$_3$/BaTiO$_3$/SrRuO$_3$                      &
          PW                                                 &
          GGA                                                &
          SrO-TiO$_2$                                        &
          3.991 \AA\ ($a^{\rm exp}_{\rm BaTiO_3}$)           &
          > 4                                                \\
                                                             &
                                                             &
                                                             &
                                                             &
          RuO$_2$-BaO                                        &
                                                             &
          > 4                                                \\
                                                             &
          Pt/BaTiO$_3$/Pt                                    &
                                                             &
                                                             &
          Pt-TiO$_2$                                         &
                                                             &
          > 4                                                \\
                                                             &
                                                             &
                                                             &
                                                             &
          Pt-BaO                                             &
                                                             &
          > 4                                                \\
                                                             &
          SrRuO$_3$/PbTiO$_3$/SrRuO$_3$                      &
                                                             &

                                                             &
          SrO-TiO$_2$                                        &
          3.905 \AA\ ($a^{\rm exp}_{\rm PbTiO_3}$)           &
          No                                                 \\
                                                             &
                                                             &
                                                             &
                                                             &
          RuO$_2$-PbO                                        &
                                                             &
          No                                                 \\
                                                             &
          Pt/PbTiO$_3$/Pt                                    &
                                                             &
                                                             &
          Pt-TiO$_2$                                         &
                                                             &
          No                                                 \\
                                                             &
                                                             &
                                                             &
                                                             &
          Pt-PbO                                             &
                                                             &
          No                                                 \\
          D. D. Fong {\it et al.} [\onlinecite{Fong-06}]     &
          SrRuO$_3$/PbTiO$_3$/vacuum                         &
          PW                                                 &
          GGA                                                &
          SrO-TiO$_2$                                        &
          $a^{\rm th}_{\rm PbTiO_3}$                         &
          $>$ 3                                              \\
                                                             &
          SrRuO$_3$/PbTiO$_3$/OH, O or H                     &
                                                             &
                                                             &
                                                             &
                                                             &
          $<$ 3                                              \\
                                                             &
          SrRuO$_3$/PbTiO$_3$/CO$_2$                         &
                                                             &
                                                             &
                                                             &
                                                             &
          $\sim$ 3                                           \\
                                                             &
          SrRuO$_3$/PbTiO$_3$/H$_2$O                         &
                                                             &
                                                             &
                                                             &
                                                             &
          $>$ 3                                              \\
          \hline
          \hline
       \end{tabular}
       }
    \end{center}
    \label{table:criticalthickness}
 \end{table}

 A summary of various results obtained with DFT based methods is
 reported in Table-\ref{table:criticalthickness}. 
 In almost all these simulations, the prototype ferroelectric
 capacitor considered is made of an insulating layer of a ferroelectric
 perovskite oxide material (typically BaTiO$_3$, PbTiO$_3$ or KNbO$_3$)
 sandwiched between metallic electrodes (typically a conductive oxide
 such as SrRuO$_3$ or a noble metal such as Pt), although lately
 a free top surface and some adsorbed molecules have been also 
 simulated.\cite{Fong-06,Spanier-06}
 Typically the capacitor is built with two symmetric metal-ferroelectric 
 interfaces, although very recently Gerra and coworkers \cite{Gerra-07} have
 explored asymmetric heterostructures and linked the first-principles
 computations with phenomenological approaches.
 The calculations were performed under short-circuit electrical boundary 
 conditions for the electrodes (zero bias), a requirement that can be 
 satisfied both in a ferroelectric/metal superlattice geometry or in a periodic 
 metal/ferroelectric/metal/vacuum geometry \cite{Kolpak-06,Kolpak-06.2} 
 (it is only recently that a method has been proposed to address finite bias
 with periodic boundary conditions \cite{Stengel-07}). In all the previous 
 approaches, 
 a monodomain configuration was considered, so the electrode was 
 the only source of screening of the polarization charge. The mechanical 
 boundary-conditions were imposed by fixing the in-plane lattice
 constant and geometry to the one imposed by the substrate. The final 
 constraint depends on the author, but the most common approach was
 to assume a strain imposed by SrTiO$_3$, the typical substrate on 
 top of which these capacitors are epitaxially grown.
 The minimum of the Kohn-Sham energies were then searched either by
 a frozen-phonon method within the ferroelectric soft mode subspace
 (line minimization displacing the atoms in the ferroelectric layer
 by hand with an amplitude corresponding to a given percentage of the 
 bulk soft mode as defined in Sec. \ref{sec:background}) or by a full atomic 
 relaxation of the atomic positions within the whole system. 
 The results  are summarized in Table-\ref{table:criticalthickness} and 
 deserve two important comments. 
 
 First, although an exhaustive comparison of the results would 
 certainly require 
 to consider also other differences in the calculations, it is observed that 
 DFT calculations performed within the 
 Generalized Gradient Approximation (GGA)  
 systematically overestimate the ferroelectric character compared 
 to those within 
 the local density approximation (LDA): while LDA predicts the 
 existence of a critical 
 thickness for ferroelectricity that ranges between two and  
 six unit cells (depending 
 on the material, electrode, interface, and strain conditions), 
 calculations within the 
 GGA often result in the absence  of such a limit, especially in 
 Pb-based perovskites, 
 with ferroelectric ground states that are stable down to the 
 ultimate thickness of 
 one unit cell, eventually with an enhancement 
 of the polarization.\cite{Sai-05,Umeno-06.1} 
 To interpret this trend it is worth 
 to notice that, although 
 GGA might a priori be considered as an improvement over LDA, it does not 
 necessarily produce more trustable results in the case of 
 ferroelectric oxides. At 
 the bulk level, the widely used GGA of Perdew, Burke and Ernzerhof (GGA-
 PBE)\cite{Perdew-96} has been shown to strongly 
 overestimate the ferroelectric 
 character and, even, to yield an erroneous super-tetragonal 
 structure in  PbTiO$_3$
 \cite{Wu-04b,Umeno-06.1,Bilc-U} and BaTiO$_3$.\cite{Bilc-U} 
 Only the recently proposed improved GGA functional of 
 Wu and Cohen \cite{Wu-06.1}  
 solves this problem and produces very accurate structures. 
 The GGA results of Table-\ref{table:criticalthickness} are 
 not making use of this new functional so that their comparison 
 with LDA results must be made  with caution : indeed, the fact 
 that GGA overestimates the ferroelectric character and concludes 
 sometimes in the absence of critical thickness might be, at least 
 partly, an artifact similar to that reported at the bulk level.
 
\begin{figure}[htbp]
 {\par\centering
  {\scalebox{0.50}{\includegraphics{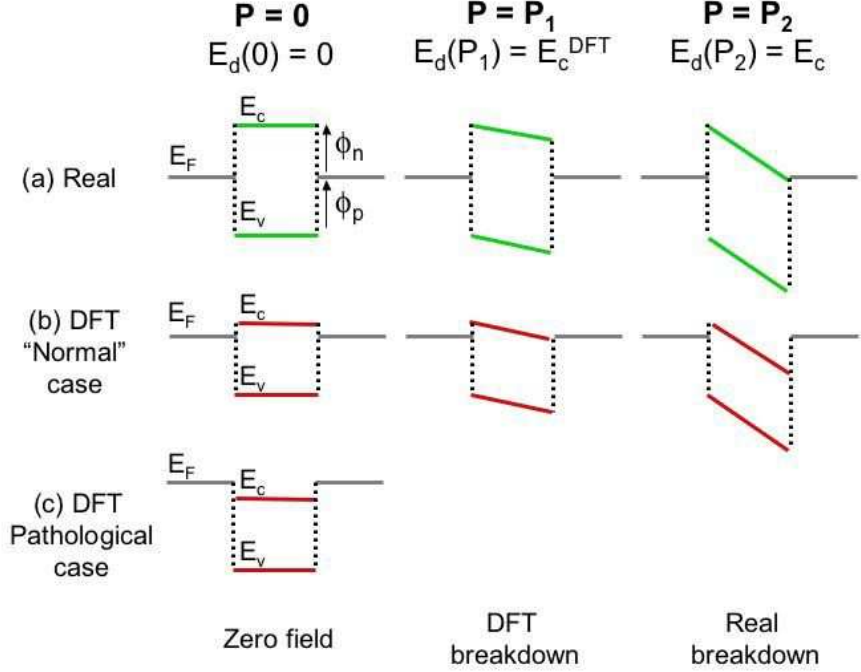}}}
 \par}
 \caption{Schematic band alignment in a typical 
          metal/ferroelectric/metal capacitor. 
          In the paraelectric state, due to the DFT bandgap problem, 
          the Schottky barrier $\phi_n$ is underestimated (b), 
          and can eventually produce pathological situations 
          where $\phi_n < 0$ (c) . 
          In the ferroelectric state ($P \neq 0$), 
          the DFT bandgap problem will additionally be responsible 
          for a lowering of the breakdown field $\mathcal{E}_c$ 
          to $\mathcal{E}_c^{\rm DFT}$. If the residual depolarizing 
          field $\mathcal{E}_d$ associated to the ground-state 
          polarization exceeds $\mathcal{E}_c^{\rm DFT}$, a spurious 
          transfer of electrons from the metal to the ferroelectric 
          conduction bands will be observed at the interface and 
          produce unphysical results.}
 \label{fig:barrier}
\end{figure}

 Second, it is well-known that the bandgap of the Kohn-Sham particles 
 as computed within DFT is significantly smaller than the experimental gap. 
 For BaTiO$_3$, typical LDA values are of the order of 1.6--1.9 eV\cite{Junquera-03.2,Bilc-U} 
 while the experimental value is of 3.2 eV. Although this DFT bandgap problem 
 does not a priori constitute a failure of the theory that should 
 internally correct for it, at least in ``exact'' DFT, in order to provide correct 
 ground-state properties, 
 it can yield unphysical results when working with approximate functionals. 
 The band alignment at a typical metal/ferroelectric interface is 
 illustrated in Fig. \ref{fig:barrier}a where, for the purpose 
 of the illustration, 
 the Fermi level of the metal is assumed to be located roughly mid-gap, 
 as typically expected from the experiment. The DFT bandgap underestimate 
 prevents accurate predictions of the barriers simultaneously for 
 electrons, $\phi_n$, and holes, $\phi_p$,  and can in many cases 
 produce pathological situations where the Fermi level of the metal 
 is erroneously aligned with the conduction bands of the ferroelectric 
 ($\phi_n$<0), producing unphysical population of the conduction bands 
 of the ferroelectric (Fig. \ref{fig:barrier}c). 
 Such a situation was, for instance, reported by the authors of 
 Ref. \onlinecite{Stengel-06} for the Pt/SrTiO$_3$ interface, 
 the structure of which had to be artificially modified to avoid the problem,
 or also appears from the analysis of the partial density of states of the 
  Pt/BaTiO$_3$ interface in Ref. \onlinecite{Velev-07}.  
 Moreover, even in ``normal'' cases where the DFT calculation reproduces 
 a small barrier for electrons in a non-polarized configuration 
 (Fig. \ref{fig:barrier}b), the situation remains problematic since the  
 underestimate of the electron barrier artificially lowers the 
 breakdown field:  if during the relaxation of the ferroelectric state, 
 the residual depolarizing field exceeds the erroneously low DFT breakdown 
 field, a spurious transfer of electrons from the metal to the ferroelectric 
 conduction bands will be observed at the interface, modifying its 
 properties and producing an unphysical relaxed structure. 
 For the SrRuO$_3$/BaTiO$_3$/SrRuO$_3$ nanocapacitor of 
 Ref. \onlinecite{Junquera-03.1}, we obtained within the LDA for the non-polar
 configuration  $\phi_p^{LDA} = 1.47$ eV and $\phi_n^{LDA} = 0.11$ eV, while 
 the use of an empirical ``scissors'' correction 
 (i.e. a rigid shift in energy of the
 BaTiO$_3$ conduction bands of 1.62 eV, to reproduce the experimental bandgap)
 provides a more realisitic value $\phi_n^{SCI} = 1.73$ eV.\cite{Junquera-03.2} 
 The artificially low value of 
 $\phi_n^{LDA}$ prevented us to perform full atomic relaxations 
 of the ferroelectric state but  was not addressed by 
 other authors reporting full relaxation on the same system.  
 
 All these issues highlight that special care must be taken when dealing 
 with metal/ferroelectric interfaces and invite to be specially 
 critical when interpreting the results. An important 
 challenge at this point for the theoriticians is to identify an alternative 
 functional avoiding the bandgap problem and its consequences on the 
 band alignment, while preserving accurate description of the structural 
 and ferroelectric properties. The B1 hybrid functional recently proposed by 
 Bilc  {\it et al.} \cite{Bilc-U} might constitute an promising option. 
   
 In spite of the technical difficulties highlighted above,  
 first-principles DFT simulations 
 remain the most accurate model currently available for the 
 theoretical study of ferroelectric capacitors and provides a wealth
 of informations
 at the atomic level. Calculations are however performed at 0 K, so that the 
 study of the temperature dependence of the structural, ferroelectric, and 
 piezoelectric properties is not accessible. Also, DFT 
 calculations are computationally very demanding and the effort required 
 to deal with such complex heterostructures, involving interfaces between 
 dissimilar metal/insulator materials, is at the limit of present capabilities. 
 To bypass these intrinsic limitations and address more complex questions, 
 alternative less accurate methods have been proposed that remain 
 based on first-principles results. These include effective 
 Hamiltonian,\cite{ZhongW-94.2,ZhongW-95} shell 
 model,\cite{Tinte-00, Tinte-01,Stachiotti-04} and molecular
 dynamics simulations combined with coarse-grained Monte Carlo 
 simulations,\cite{Shin-07}
 where the different parameters are directly fitted on DFT results.

 Wu and coworkers \cite{Wu-04a,Wu-05b} have performed effective Hamiltonian
 simulations on Pb(Zr$_{0.5}$Ti$_{0.5}$)O$_3$ free standing slabs.
 The choice of the Ti and Zr composition is interesting, since it is close
 to the morphotrophic phase boundary of bulk PZT.
 Since the calculations were done in the absence of surface charge screening,
 films thicker than 4 unit cell form 180$^\circ$ domains in the direction
 normal to the surface to screen the depolarizing field,
 so the average component of the out-of-plane local mode vanishes, independently
 of the strain.
 The depolarizing field does not influence the in-plane polarization versus 
 strain behaviour.
 A rich strain-temperature phase diagram, including a monoclinic and a 
 stripe domain phase not present in the bulk parent material was predicted.

 For the same composition of Pb(Zr$_{0.5}$Ti$_{0.5}$)O$_3$, Kornev
 {\it et al.} \cite{Kornev-04} extended the effective Hamiltonian 
 in order to incorporate explicitly the presence of a free surface
 and the effects of a partial compensation 
 of the depolarizing field. 
 As in Ref. \onlinecite{Nagarajan-06}, in this first attempt, only the net 
 component of the out-of-plane polarization at the 
 surface layers was considered to compute the 
 homogeneous depolarizing field and the corresponding contributions to
 the energy.
 The treatment of $\mathcal{E}_{d}$ was then refined by Ponomareva
 and coworkers,\cite{Ponomareva-05.2} who corrected the dipole-dipole
 interactions of the bulk model hamiltonian by summing explicitely 
 contributions of the local dipoles present within the film,  
 and derived  in this way 
 an {\it exact} expression for the depolarizing energy and field
 at an atomistic level in any low dimensional ferroelectric structure
 under open circuit (OC) boundary conditions. 
 If the polarization lies along a non-periodic direction, 
 the OC boundary conditions naturally lead to the existence 
 of a maximum of the depolarizing field that is later screened by hand.
 The amount of screening and therefore the amplitude of the residual field
 can be tuned with a parameter, $\beta$ (see Section III-B), that plays a 
 role similar to 
 the effective screening length in the model described above
 and ranges between $\beta$ = 0 (OC boundary conditions) and 
 $\beta$ = 1 (perfect short circuit boundary conditions).
 This model has been applied 
 to different kinds of low dimensional ferroelectrics.\cite{Ponomareva-05.3}
 
 In Ref. \onlinecite{Kornev-04}, the interplay between the 
 screening of the depolarizing field and the strain imposed by
 the substrate was also explored.
 In the limit of perfect screening, a uniform
 polarization along $z$ is predicted, while the behaviour of the
 in-plane components of the polarization depends on the strain,
 following the general rule described in Sec. \ref{sec:strain}:
 a compressive strain supresses the in-plane polarization while a
 tensile strain favours the presence of a $aa$-phase.
 Therefore, the phase change from a monoclinic phase (the so-called
 $M_{A}$ phase with
 $\langle P_{x} \rangle = \langle P_{y} \rangle \ne 0$,
 $\langle P_{z} \rangle > \langle P_{x} \rangle$) for
 sufficiently large tensile strains to a tetragonal phase with
 $\langle P_{x} \rangle = \langle P_{y} \rangle = 0$,
 $\langle P_{z} \rangle > 0$ for stress free or compressive strains.
 On the other extreme case of open-circuit boundary conditions,
 the {\it average} polarization along $z$ vanishes
 independently of strain, 
 while the in-plane polarization aligns with the [010] direction
 ($\langle P_{y} \rangle \ne 0, \langle P_{x} \rangle = 0$)
 for the stress free, displays a monoclinic $M_{c}$ phase 
 ($\langle P_{x} \rangle \ne \langle P_{y} \rangle \ne 0$,
 $\langle P_{y} \rangle > \langle P_{x} \rangle$) under tensile strain,
 or vanishes for compressive strain.
 For intermediate values of the depolarizing field,
 the polarization continuously rotates and passes through low-symmetry phases.
 Interestingly, these phases are stable only for compositions lying near the 
 morphotrophic phase boundary of bulk PZT, disappearing 
 in favour of a tetragonal phase for high enough Ti-concentrations.
 The fact that the {\it average} out-of-plane polarization vanishes
 under open circuit boundary conditions
 does not mean that the local dipoles along $z$ vanishes.
 On the contrary, they might have a large value, close to the bulk one,
 but forming domains.
 The process of formation of these domains was monitored with respect
 the screening of the depolarizing field.
 When $\mathcal{E}_{d}$ is large enough, some ``bubbles'' 
 (cylindric nanodomains having local dipoles that are aligned in an opposite
 direction with respect the one in the monodomain configuration)
 nucleate and grow laterally till 180$^\circ$ stripe domains are formed.

 Model hamiltonian techniques have the additional advantage of a trivial
 treatment of external electric fields.
 Starting from a five unit cell thick thin film geometry under
 compressive strain, where $c$-oriented 180$^\circ$ stripe domains are
 stabilized even assuming a realistic amount of screening of the
 depolarizing field, the evolution of the domain structure under an electric
 field has been studied for Pb(Zr$_{0.5}$Ti$_{0.5}$)O$_{3}$ \cite{Lai-06}
 and BaTiO$_{3}$,\cite{Lai-07} giving some light into the
 mechanisms to saturate the films.
 The path is essentially the same for both perovskites oxides and consist
 of three different steps: $(i)$ the growth of the domain parallel to
 the field at the expense of the other up to a first threshold value
 of the external field; $(ii)$ the formation of the nanobubbles 
 defined before; and $(iii)$ the contraction of the nanobubbles till 
 the final collapse into a monodomain configuration at a second
 critical value of the external field.
 The main differences between BaTiO$_{3}$ and Pb(Zr$_{0.5}$Ti$_{0.5}$)O$_{3}$
 are $(i)$ the orientation of the domain walls: the stripes alternate
 along [110] for BaTiO$_{3}$ rather than 
 along [100] for Pb(Zr$_{0.5}$Ti$_{0.5}$)O$_{3}$, and
 $(ii)$ the ``hardness'' of the polarizarion in BaTiO$_3$ to rotate
 (BaTiO$_{3}$ films profoundly dislike significantly 
 rotating and in-plane dipoles).

 The formation and response of the ``bubbles'' to external electric fields
 and the recent predictions of 
 domains of closure \cite{Landau-35,Kittel-46,Kittel-49} 
 (very common in ferromagnets but detected in ferroelectric thin films
 only recently \cite{Scott-U}), 
 have bridged 
 a fascinating parallelism between ferroelectric and ferromagnetic domains.
 Although the shape and size of the ``bubbles'' (elliptical in ferroelectrics
 and spherical in ferromagnets), the abruptness of the
 domain walls (one lattice constant thick in ferroelectrics and 
 much more gradual, over many atomic planes, in ferromagnets),
 or the exceptionally small periodicity and constancy with the field of the
 ferroelectric domain periods differ from the counterpart ferromagnetic
 properties, they might not be so different as traditionally considered,
 despite the profound differences between electrostatic and magnetostatic
 interactions.
 Domains of closure have been predicted 
 using a first-principles
 effective hamiltonian for Pb(Zr$_{0.4}$Ti$_{0.6}$)O$_3$
 \cite{Prosandeev-07} asymmetrically screened
 (grown on a nonconducting substrate and with a metal with a dead layer
 as top electrode),
 and using a Landau-Ginzburg phenomenological approach for a PbTiO$_3$ thin
 film,\cite{Stephenson-06} both asymmetrically and symmetrically coated
 with insulating SrTiO$_3$.
 These predictions have obtained further credit after full
 first-principles simulations on SrRuO$_{3}$/BaTiO$_{3}$/SrRuO$_{3}$
 ferroelectric capacitors by Aguado-Puente and Junquera,\cite{Aguado-Puente}  
 where the domains of closure are obtained
 even for a symmetrical metal/ferroelectric/metal capacitor,
 where the metallic plates should provide significant screening.
 As it happens with ferromagnetic domains, ferroelectric domains 
 follow the Kittel law 
 (that states that the domain width is directly
 proportional to the square root of the sample's thickness), 
 at least for thicknesses thicker than 1.2 nm.\cite{Lai-07.2}

 The effective Hamiltonian approach is now considered as a standard method and 
 has been used to address various other questions. The phase diagram of PZT 
 films as a function of temperature and Ti composition under open-circuit 
 boundary conditions has been reported in Ref. \onlinecite{Almahmoud-04}.  
 Other systems like BaTiO$_3$ have also been considered for which the 
 strain/temperature phase diagram has been obtained and the effect of 
 incomplete screening of $\mathcal{E}_d$ has been discussed.\cite{Paul-07}
 Although most studies focused on [001] epitaxial films, the influence 
 of the growth direction on the properties was also 
 investigated.\cite{Ponomareva-06} 

 Besides all the previous theoretical frameworks,
 molecular dynamics combined with coarse-grained Monte Carlo simulations
 have been recently used to study a very important tecnological problem:
 the swithching of the ferroelectric polarization.\cite{Shin-07}
 These simulations show that the shape of the critical nucleus is square and
 not a thin triangular plate as suggested by the prevailing
 Miller-Weinrich model. 
 The critical nuclei grow in an anysotropic way, with activation barriers
 for sideways and forward growth much smaller than for nucleation.
 The values for the activation energies and activation fields 
 predicted by the molecular dynamics simulations are one order of
 magnitude smaller than those expected from the Miller-Weinrich theory,
 mainly due to the new slanted interface model for the surface of the nuclei,
 and the domain velocities are in very good agreement with the
 ones measured experimentally.\cite{Tybell-02}

\section{Superlattices}
\label{sec:superlattices}

 Artificial ferroelectric superlattices are nowadays considered 
 as an interesting alternative to ferroelectric thin films \cite{Rijnders-05} 
 since, as it is discussed below, they allow fine tuning of the ferroelectric 
 properties while maintaining prefect crystal ordering. The 
 large variety of materials that can be combined (ferroelectrics, 
 incipient ferroelectrics or regular insulators) and the way these materials 
 can be ordered  within the structure (bicolor and tricolor superlattices 
 or even more complex structures with composition gradients) also offers 
 tremedous scope for creating artificial ferroelectric materials with possibly
 new and tailor-made properties. 

 During the recent years, short-period epitaxial multilayers combining a 
 ferroelectric and an incipient ferroelectric material 
 (like BaTiO$_3$/SrTiO$_3$,\cite{Iijima-92,Tabata-94,Tabata-97,Zhao-99,
 Tsurumi-02,Shimuta-02,Jiang-03,Rios-03,Neaton-03,Johnston-05,Lee-06,
 Tenne-06,Lisenkov-07,Tian-06} 
 PbTiO$_3$/SrTiO$_3$,\cite{Jiang-99,Dawber-05.2,Dawber-U,Bousquet-U}  
 KNbO$_3$/KTaO$_3$, 
 \cite{Specht-98,Sepliarsky-01.1,Sepliarsky-01.2,Sepliarsky-02}
 and related tricolor systems like 
 BaTiO$_3$/SrTiO$_3$/CaTiO$_3$ \cite{Lee-05,Nakhmanson-05,Nakhmanson-06})
 have retained most of the attention both at the theoretical and experimental 
 level.  In these systems, proper handling of the mechanical and 
 electrical boundary conditions appears crucial to monitor the 
 ferroelectric properties in a way very similar to what was 
 previously discussed for thin films.

 \begin{figure}[htbp]
 {\par\centering
  {\scalebox{0.70}{\includegraphics{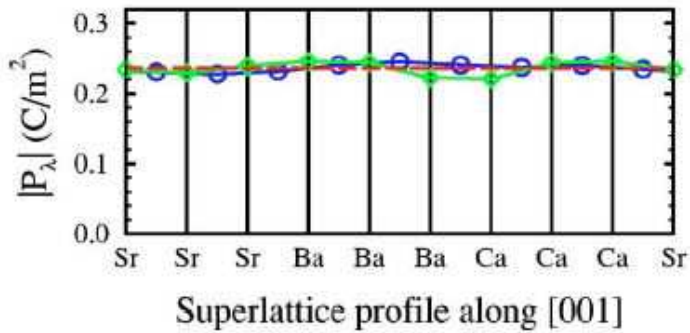}}}
 \par}
  \caption{Layer by layer polarization profile along a ferroelectric 
         BaTiO$_3$/SrTiO$_3$/CaTiO$_3$ multilayer as obtained from DFT 
         first-principles calcualtions. A polarization similar to that of 
         BaTiO$_3$ is induced in the non-ferroelectric 
         SrTiO$_3$ and CaTiO$_3$ layers. Figure taken 
         from Ref. \onlinecite{Nakhmanson-05}.}
  \label{fig:superlattice}
\end{figure}

 First, the alternation of ultrathin layers is particularly suited 
 to impose coherent epitaxial strains and achieve ``strain engineering'' 
 of the properties. The compressive epitaxial strain produced by a 
 SrTiO$_3$ substrate on BaTiO$_3$ layers contributes to explain 
 polarization enhancement with respect to the bulk BaTiO$_3$ value 
 in BaTiO$_3$/SrTiO$_3$ bicolor superlattices \cite{Neaton-03,Tian-06,Tenne-06} 
 and even in tricolor \cite{Lee-05} superlattices, despite the fact that part 
 of the system is not nominally ferroelectric.  
 The epitaxial strain additionally produces significant shift of the phase 
 transition temperature as for instance reported for 
 BaTiO$_3$/SrTiO$_3$  \cite{Tenne-06} 
 and PbTiO$_3$/SrTiO$_3$ \cite{Dawber-U}  superlattices.
 Ferroelectricity is even expected in SrZrO$_3$/SrTiO$_3$ superlattices 
 on SrTiO$_3$, in spite of the fact that neither SrZrO$_3$ nor SrTiO$_3$ is 
 ferroelectric at the bulk level.\cite{Tsurumi-04} Although 
 it was not stated by the 
 authors, this can probably be partly explained from the large compressive 
 strain imposed on the SrZrO$_3$ layers by the SrTiO$_3$ substrate.

 Second, electrostatic coupling between the layers is also playing a key role. 
 On the one hand, polarizing the ferroelectric BaTiO$_3$ layers only in a 
 BaTiO$_3$/SrTiO$_3$ superlattice would produce a large polarization 
 gradient at the interfaces, inducing huge depolarizing fields and 
 electrostatic energy. On the other hand, polarizing SrTiO$_3$ also 
 has an energy cost since this material is not spontaneously polarized. 
 The ground-state of the system will therefore depend on the 
 competition between different energy terms and can strongly evolves 
 depending of the compounds involved and the superlattice period. 

 In short-period superlattices where the ferroelectric alternates 
 with a highly polarizable compound such as an incipient ferroelectric 
 (like SrTiO$_3$ or KTaO$_3$), a polarization will be induced in the 
 latter in order to avoid  depolarizing fields and the superlattice 
 will typically exhibit a homogeneously polarized configuration 
 (see Fig. \ref{fig:superlattice}). First pointed out by Neaton and Rabe on 
 BaTiO$_3$/SrTiO$_3$ superlattices, such a behavior was also 
 theoretically predicted for PbTiO$_3$/SrTiO$_3$,\cite{Dawber-05.2}
 or even BaTiO$_3$/SrTiO$_3$/CaTiO$_3$ \cite{Nakhmanson-05,Nakhmanson-06} 
 short-period superlattices.  In many theoretical studies, the layer 
 polarization was estimated roughly, multiplying the atomic displacements
 into the layer by the bulk Born effective charges.
 A more rigorous and accurate approach has been proposed recently 
 based on Wannier functions.\cite{Wu-06.3}

 In a recent work, Dawber {\it et al.} \cite{Dawber-U} have 
 demonstrated that key ferroelectric properties can be tuned over a 
 very wide range in PbTiO$_{3}$/SrTiO$_{3}$ superlattices:  it was shown 
 that the polarization can be adjusted from 0 to 60 $\mu {\rm C/cm^2}$ and 
 the transition temperature from room temperature to 700$^\circ$C 
 by properly adjusting the volume fraction of PbTiO$_3$.  
 Moreover, a simple Landau-based model was proposed that allows to 
 determine a priori the properties of the system, so opening the way 
 to the production of samples with ferroelectric properties designed 
 for particular applications. It was also reported that the presence of 
 SrTiO$_3$ between the PbTiO$_3$ layers in the superlattice improve the 
 electrical properties and allows to avoid the large leakage currents 
 typically reported in PbTiO$_3$ thin films while maintaining 
 perfect crystal structure. 

 The simple arguments described above have been shown to provide 
 reasonable description of (PbTiO$_3$)$_n$/(SrTiO$_3$)$_3$ superlattices 
 over a relatively wide range of PbTiO$_3$ thicknesses 
 (for $n$ ranging from 1 to 53).\cite{Dawber-05.2}
 They are however expected to apply mainly to relatively short-period 
 systems.  In longer period superlattices, other behaviors are expected. 
 In KNbO$_3$/KTaO$_3$ \cite{Sepliarsky-01.1} it was theoretically 
 shown using a shell-model that the electrostatic coupling between 
 layers (responsible for a homogeneous polarization throughout the system) 
 will progressively decrease as the period increases, yielding a progressive 
 reduction of KTaO$_3$ polarization. In BaTiO$_3$/SrTiO$_3$,\cite{Lisenkov-07}
 effective Hamiltonain simulations have confirmed that short-period 
 superlattices behave like a homogeneous-like material with a phase 
 diagram that resembles that of a (001) BaTiO$_3$ thin film under 
 short-circuit conditions while inhomogeneous atomic features and 
 original domain patterns were predicted in longer period systems. 
 Appearance of domain structures in ferroelectric/paraelectric superlattices 
 was also discussed by Stephanovich {\it et al.}\cite{Stephanovich-05}

 The previous discussion implicitly focused on the evolution of the 
 out-of-plane polarization.  In some cases, an in-plane component 
 of the polarization can also eventually appear. This was predicted in 
 KNbO$_3$/KTaO$_3$ superlattices in which the epitaxial compressive
 strain is not strong enough to force KNbO$_3$ (with a rhombohedral
 ground-state) to become tetragonal.\cite{Sepliarsky-02} In a way compatible  
 with a symmetry lowering observed experimentally,\cite{Rios-03} an 
 in-plane polarization component was also predicted in the SrTiO$_3$ layers 
 of BaTiO$_3$/SrTiO$_3$ superlattices under 
 in-plane expansion.\cite{Johnston-05} 
 The main difference between the out-of-plane and in-plane polarization 
 components is that while the out-of-plane components of the different 
 layers are strongly electrostatically coupled as previously discussed, 
 the in-plane components behaves much more independently and remain 
 confined in the individual layers in which they are induced by the epitaxial 
 strain conditions imposed by the substrate. 
  
 The interest for ferroelectric superlattices is not limited to the fact 
 that they allow the tuning of the ferroelectric properties as discussed 
 above but also results from their potentiality to generate totally new 
 behaviors or phenomena. In Pb(Sc$_{0.5+x}$Nb$_{0.5-x}$)O$_{3}$ 
 (PSN) \cite{George-01} and Pb(Zr$_{1-x}$Ti$_{x}$)O$_{3}$ (PZT) \cite{Huang-03} 
 superlattices, it is expected that symmetry lowering can yield new 
 stable polar phases and strong enhancement of the functional properties. 
 Unusual thermoelectric properties and nonergodicity were also predicted 
 in PZT superlattices.\cite{Kornev-03} In very short-period 
 PbTiO$_3$/SrTiO$_3$  
 superlattices an unexpected recovery of ferroelectricity has been observed in 
 the limit of ultrathin PbTiO$_3$ layers that cannot be 
 explained from the simple 
 arguments reported above.\cite{Dawber-05.2,Dawber-U, Bousquet-U}. 
 Beyond such bicolor systems, tricolor 
 superlattices, which break the inversion 
 symmetry \cite{Sai-00} are also expected 
 to potetially exhibit unusual and attractive properties. 
 
 First-principles based simulations contributed to make significant progresses 
 in the understanding of the behavior of ferroelectric superlattices and even 
 opened the door to a possible theoretical design of artificial ferroelectric 
 nanostructures with predetermined and 
 tailor-made properties.\cite{Iniguez-01.1}

\section{Nanoparticles and Nanowires}
\label{sec:nanoparticles}

 Although most of the recent experimental and theoretical efforts 
 on ferroelectric
 nanostructures concerned thin films geometries, the growth and 
 characterization 
 of ferroelectric nanowires,\cite{Yun-02, Wang-06.2}
 nanotubes,\cite{Luo-03,Morrison-03.2} nanocolumns, \cite{Schilling-06}
 nanoislands,\cite{Chu-04} or even ``nanosolenoids'' \cite{Zhu-06}
 are at an incipient and exciting state. The geometries of these 
 nanostructures favor novel inhomogeneous polarization and strain 
 configurations, strongly dependent of the electrical and mechanical
 boundary conditions.
  
 DFT first-principles studies of such low dimensional structures are, 
 however, computationally very demanding.  
 Other recent studies made use of first-principles based 
 shell-models \cite{Stachiotti-04} and model hamiltonian methods.\cite{Fu-03}
 For instance, domain structures, polarization and coercive fields of 
 nanoscopic particles of BaTiO$_3$ have been carried out using such
 methodologies. 

 \begin{figure}[htbp]
 {\par\centering
  {\scalebox{0.80}{\includegraphics{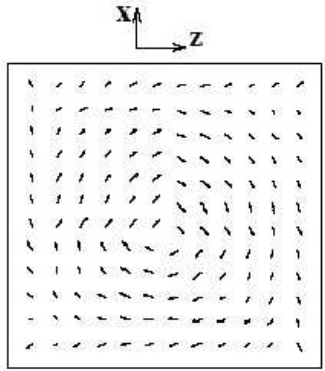}}}
 \par}
  \caption{``Vortex-like'' structure of the local polarization in a 
          $( 12 \times 12 \times 12)$ BaTiO$_3$ quantum dot.
          The figure corresponds to a cut along the $y$ = 6$^{th}$ plane.
          Although the macroscopic polarization vanishes
          the toroid moment of the vortex structure has a well-defined 
          finite value. Figure taken from Ref. \onlinecite{Fu-03}.}
  \label{fig:dot}
\end{figure}

 Perhaps, the most remarkable result is the prediction of novel
 exotic geometric ordering of ferroelectricity in BaTiO$_3$ \cite{Fu-03} and
 Pb(Zr$_{0.5}$,Ti$_{0.5}$)O$_3$ \cite{Naumov-04,Scott-05} 
 nanoscale disks and rods, where the  
 the local polarization 
 rotate from cell to cell, forming a ``vortex-like'' pattern.
 Although the macroscopic polarization of these quantum dots vanishes at
 zero applied field,
 they exhibit a new order parameter, the {\it toroid moment} of polarization, 
 introduced by Naumov {\it et al.} \cite{Naumov-04} and defined as

 \begin{equation}
    \vec{G} = \frac{1}{2N} \sum_{i} \vec{R}_{i} \times \vec{p}_{i},
    \label{eq:toroid}
 \end{equation}

 \noindent where $\vec{p}_{i}$ is the local dipole of cell $i$ located at
 $\vec{R}_{i}$, and $N$ is the number of unit cells in the simulation.
 Since the local dipoles in the vortex can rotate either clockwise
 or anticlockwise, the toroid moment might adopt two different
 values pointing in opposite directions, so that one bit of information
 might be stored by assigning one value of the Boolean algebra (``1'' or ``0'')
 to each of these states.
 The direction of the toroidal moment can be efficiently controlled
 by a transverse inhomogeneous electric field generated by two opposite
 charges located away from the studied 
 dot,\cite{Prosandeev-06.1} opening exciting oportunities for 
 nanomemory devices.
 Technologically, the ferroelectric vortex phase promises to increase
 the storage density of ferroelectric random access memories (Fe-RAM) by
 five orders of magnitude.
 
 In order to investigate the possibility of engineering the dipole patterns
 and vortex structures, a study of 
 the phase diagram as a function of temperature of these ferroelectic 
 nanodots when they are embedded in a polarizable medium 
 was reported by Prosandeev and Bellaiche.\cite{Prosandeev-06.2}
 Six different phases were found, depending on the ferroelectric strength
 of the material constituting the dot and of the system forming the
 medium. Some of the novel phases are remarkable, for instance
 the coexistence of a toroidal moment and a spontaneous polarization
 at low T when a soft ferroelectric dot is immersed in a medium that
 is ferroeletrically harder than the dot.
 Medium driven interactions between dots were demonstrated with adjacent
 vortices rotating in an opposite fashion (``antiferrotoroidic'' phase).

 The phase transition from the vortex structure (with the local dipoles
 circularly ordered) to a ferroelectric phase (with all the local
 dipoles pointing along the same direction and with an out-of-plane 
 polarization) in a Pb(Zr$_{0.5}$Ti$_{0.5}$)O$_{3}$ nanocylinder
 by an electric field perpendicular to the
 plane of the vortex has been recently addressed.\cite{Naumov-07}
 The transition between the two phases is through an intermediate state 
 that generates a lateral toroid moment and breaks the macroscopic cylindrical
 symmetry. The observation of this new phase yielded to the discovery 
 of a stricking collective phenomenon, that leads to the 
 annihilation of the ferroelectric vortex in a peculiar azimuthal mode.

 A first step to explain the existence of this toroidal moment 
 in ferroelectric nanoparticles from a DGL theory has been taken 
 by Wang and Zhang \cite{Wang-06}
 A more exhaustive review of all these computations can be found
 in Ref. \onlinecite{Ponomareva-05}.

 In between the two-dimensional (2D) thin films and superlattices 
 discussed in Secs. \ref{sec:films} and 
 \ref{sec:superlattices}, and the zero-dimensional (0D) nanoparticles
 described above, the case of one-dimensional (1D) nanowires was
 also recently addressed.
 Model hamiltonian simulations on Pb(Zr,Ti)O$_{3}$ \cite{Naumov-05} 
 and full first-principles
 simulations on BaTiO$_{3}$ nanowires \cite{Geneste-06}
 have demonstrated how long-range ordering 
 exists in these 1D systems even at finite temperature and under zero 
 external field. 
 These results challenged traditional statistical models that claimed
 that 1D lattices with particle-particle interactions decaying faster than
 $r^{-n}, n \ge 3,$ are impossible to have phase transitions at finite T.
 As it happened in thin films and nanoparticles, the mechanical 
 and electrical boundary conditions are essential in the nanowires.
 The presence of uncoated free surfaces yields to the existence of 
 depolarizing fields along two directions and forces the collective polarization
 to lye along the longitudinal direction.
 A critical diamenter for the polarization was estimated around 1.2 nm 
 for BaTiO$_{3}$ \cite{Geneste-06} and 2.0 nm for 
 Pb(Zr,Ti)O$_{3}$,\cite{Naumov-05} below which the ferroelectricity 
 was suppresed. At smaller radii, low atomic coordinations at the surface
 produce a contraction of the unit cell that is responsible
 for the suppression of the ferroelectric distortion at the critical radius.
 Nevertheless, even below this threshold size of the diameter,
 ferroelectricity might be recovered playing with the mechanical boundary
 conditions, by applying appropriate tensile strain.

 Previous first-principles simulations on nanowires also contributed
 to get a new insight into the concept of the ferroelectric correlation 
 volume, defined as the smallest volume within which atomic displacements
 must be correlated in order to decrease the energy of the crystal
 and produce a stable polar entity. Analyzing the critical diameter of the 
 wire as a function of its length, the computations prove that is
 the anisotropic shape of the polar domain rather than its volume that
 determines the stability of its ferroelectric state.\cite{Geneste-U}

\section{Conclusions and perspectives}

 As illustrated throughout this review, various progresses have been
 reported during the last years concerning the description of the
 ferroelectric properties in oxide ultrathin films and superlattices.
 In this context, first-principles theory and modeling constituted an
 efficient guide and provided a significant support to experimental
 works. On the one hand, proper understanding of the crucial role
 played by the electrical and mechanical allowed to realize significant
 advances as discussed above. On the other hand, however,
 many issues still remain unclear so that this field of research is still
 at an exciting and incipient stage.
 
 In ferroelectric thin films, despite recent achievements, some important
 questions remain  without clear answer and interesting perspectives ask
 for further confirmations. Only recently addressed from 
 first-principles,\cite{Spanier-06} the mechanism and the 
 quality of the screening at
 free surfaces with air remains unclear.
 Also the reason why some ferroelectric films on metallic electrodes stay
 monodomain with a reduced polarization while others break into domains
 is particularly intriguing.\cite{Lichtensteiger-07} The roles of defects, such
 as oxygen vacancies, and of leakage currents on the amplitude of the
 depolarizing field was up to now neglected in theoretical studies while
 they might play a role, at least in some cases. The mechanism of the
 switching and domain-wall motion in ultrathin films is an important question
 that has been studied in detail experimentally
 \cite{Tybell-02,Paruch-05,Paruch-06,Paruch-07} but that has not been
 addressed yet at the first-principles level,
 although the first-steps on multi-scale computations based
 on molecular dynamics and coarse-grained Monte Carlo simulations
 have been recently taken.\cite{Shin-07}
 Recent studies reported interesting perspectives for ferroelectric tunnel
 junctions \cite{Zhuravlev-05,Tsymbal-06,Velev-07} but no first-principles 
 quantitative estimate of the tunneling current and of the electroresistance 
 has been reported yet.  
 Finally and without being exhaustive, the case of a ferroelectric
 layer between magnetic electrodes was proposed as a mechanism to
 achieve ferroelectric control of magnetism \cite{Duan-06.3,Sahoo-07} 
 but is still widely unexplored.

 Ferroelectric superlattices constitute a promising alternative avenue that
 offers practical solutions  to bypass some of the drawbacks of epitaxial
 ultrathin films (huge depolarizing fields, high leakage currents, strain
 relaxation), while preserving perfect crystal ordering and allowing to monitor
 the ferroelectric properties. Up to now, fine tuning of the polarization and
 phase transition temperature has, for instance, been illustrated playing with
 the superlattice period and composition but the most exciting 
 perspectives probably
 rely on the potentiality of such systems to generate totally new and unexpected
 phenomena. 
 Also, superlattices might offer particularly attractive opportunities in the
 emergent field of magneto-electric multiferroics.\cite{Hatt-07}

 The perspectives are numerous and first-principles simulations will certainly
 continue to play an active role in the field of nanoscale ferroelectrics 
 in the future. In some cases, this will mandatory require new theoretical 
 developments such as, for instance, the identification of alternative and more 
 accurate functionals or of new efficient approaches to compute 
 complex properties. 
 In order cases, the computer power will constitute the 
 main limitation. In all cases, pursuing the constructive 
 dialog initiated during the 
 recent years between theorists and experimentalists will 
 certainly appear as a major motivation.

\section*{Acknowledgements}
 The authors thank J.-M. Triscone, M. Dawber, C. Lichtensteiger, 
 H. Kohlstedt, M. Alexe, G. Geneste, E. Bousquet,
 and P. Aguado-Puente for valuable discussions.
 PhG acknowledges financial support
 from the VolkswagenStiftung (I/77 737), the Interuniversity Attraction 
 Poles Program - Belgian State - Belgian Science policy (P6/42), 
 the  FAME European Network of Excellence and the 
 MaCoMuFi European Strep project.
 JJ acknowledges financial support by the Spanish 
 MEC under Project FIS2006-02261, and the Australian 
 Research Council ARC Discovery Grant DP 0666231.

\end{document}